\documentclass[twocolumn]{aastex631}

\usepackage{amsmath}
\usepackage{mathtools}
\usepackage{url}
\usepackage{soul}
\shorttitle{LSST Interstellar Objects}
\shortauthors{Mar{\v{c}}eta \& Seligman}

\begin{document}

\title{ Synthetic Detections of Interstellar Objects with The Rubin Observatory Legacy Survey of Space and Time } 
 \author[0000-0003-4706-4602]{Du{\v{s}}an Mar{\v{c}}eta }
 \affiliation{Department of Astronomy, Faculty of Mathematics, University of Belgrade, Studentski trg 16, Belgrade, 11000, Serbia}

 \correspondingauthor{Du{\v{s}}an Mar{\v{c}}eta }
 \email{dusan.marceta@matf.bg.ac.rs}

 \author[0000-0002-0726-6480]{Darryl Z. Seligman}
 \altaffiliation{NSF Astronomy and Astrophysics Postdoctoral Fellow}
 \affiliation{Department of Astronomy and Carl Sagan Institute, Cornell University, 122 Sciences Drive, Ithaca, NY, 14853, USA}

\begin{abstract}
The discovery of two interstellar objects passing through the Solar System, 1I/`Oumuamua and 2I/Borisov, implies that a galactic population exists with a spatial number density of order $\sim0.1$ au$^{-3}$. The forthcoming Rubin Observatory Legacy Survey of Space and Time (LSST) has been predicted to detect more asteroidal interstellar objects like 1I/`Oumuamua. We apply recently developed methods to simulate a suite of galactic populations of interstellar objects with a range of assumed kinematics, albedos and size-frequency distributions (SFD).  We  incorporate these populations into the objectsInField (OIF) algorithm, which simulates detections of moving objects by an arbitrary survey.  We find that the LSST should detect between $\sim 0-70$ asteroidal interstellar objects every year (assuming the implied number density), with sensitive dependence on the SFD slope and characteristic albedo of the host population. The apparent rate of motion on the sky --- along with the associated trailing loss --- appears to be the largest barrier to detecting interstellar objects. Specifically, a relatively large  number of synthetic objects  would be detectable by the LSST if not for their rapid sky-motion ($>0.5^\circ$ d$^{-1}$). Therefore, algorithms that could successfully link and detect rapidly moving objects would significantly increase the number  of interstellar object discoveries with the LSST (and in general).  The mean diameter of detectable, inactive interstellar objects ranges from $\sim50 - 600$ m and depends sensitively  on the SFD slope and albedo.
\end{abstract}


\section{Introduction} \label{sec:intro}

The source reservoirs of small bodies within the Solar System have long been the subject of investigation. The isotropically distributed in inclination  long-period comets (LPCs) stem from the spherical Oort cloud \citep{Oort1950} which has a total mass of $\sim1-20 $ M$_\oplus$ \citep{francis2005,Kaib09,Brasser13,Dones15}. Meanwhile, the low inclination short-period comets (SPCs) come from the Kuiper belt Objects (KBOs) \citep{Jewitt1993}. The Kuiper belt and Oort cloud were presumably populated during  significant and  early orbital migration  of the giant planets \citep{Hahn1999,Gomes2004,Tsiganis2005,Morbidelli2005,Nesvorny2018}. However,  only a small fraction,  $\sim1-10\%$, of scattered objects  populated the Kuiper belt and Oort cloud \citep{Hahn1999,Brasser10,Dones15,Higuchi15}. The Solar System  likely generated $\sim1-30$  M$_\oplus$ of material in interstellar comets \citep{Seligman2023}.

The discovery of the first interstellar object 1I/`Oumuamua implies that  a galactic source population exists with a spatial number density of order  $\sim n_{o}\sim1\times 10^{-1}\,$\textsc{au}$^{-3}$ \citep{Trilling2017,Laughlin2017,Do2018,Levine2021}. Extrapolation of this spatial number density to an isotropic galactic  population of similar objects implies that on average $\sim1$ M$_\oplus$ of material is ejected by every stellar system \citep{Jewitt2022_ARAA}. Therefore,  the discovery of future interstellar interlopers appears to be imminent.

For recent reviews of this field, we refer the reader to \citet{Jewitt2022_ARAA}, \citet{MoroMartin2022}, \citet{Fitzsimmons2023} and \citet{Seligman2023}.  1I/`Oumuamua  was discovered on October 19 2017 \citep{Williams17} with the Pan-STARRS telescope \citep{Chambers2016}. It lacked a cometary tail \citep{Meech2017,Jewitt2017,Ye2017} and had a slightly reddened reflectance spectra \citep{Bannister2017, Masiero2017, Fitzsimmons2017, Bolin2017}, an elongated shape \citep{Meech2017,Bannister2017,Jewitt2017,Knight2017,Bolin2017,Drahus18,Fraser2017,Belton2018,Mashchenko2019}, a low incoming velocity with respect to the Local Standard of Rest \citep{mamajek2017, Gaidos2017a,Hallatt2020,Hsieh2021} and a nongravitational acceleration \citep{Micheli2018}. It has been hypothesized that 1I/`Oumuamua is a porous fractal aggregate \citep{moro2019fractal,Sekanina2019b,Flekkoy19,Luu20} or a sublimating icy comet with little dust production \citep{fuglistaler2018,Sekanina2019,Seligman2020,Levine2021,Levine2021_h2,desch20211i,jackson20211i,Desch2022,Bergner2023}.

A second interstellar object, 2I/Borisov was discovered 2 years later in 2019. The object had a nuclear radius estimated to be 0.2--0.5 km \citep{Jewitt2019b,Jewitt20}, and a distinct coma \citep{Jewitt2019b,deleon2019,Bolin2019,Guzik:2020,Hui2020,Mazzotta21}. The object had some nontypical properties including a hypervolatile enriched composition \citep{Cordiner2020,Bodewits2020}, high polarization of dust in the outflow  \citep{Bagnulo2021,Halder2023} and a disintegration event \citep{Drahus2020ATel,Jewitt2020:BorisovBreakup, Jewitt2020ATel, Bolin2020ATel,Zhang2020ATel,Kim2020}

There has been significant efforts to characterize  the prospects for detecting  interstellar objects in the Solar System. The forthcoming Rubin Observatory Legacy Survey of Space and Time (LSST)  \citep{jones2009lsst,Ivezic2019} should efficiently detect transient objects \citep{solontoi2011comet,Veres2017,veres2017b,Jones2018}. Moreover, the forthcoming NEO Surveyor \citep{Mainzer2015} should also detect interstellar objects. There has been significant effort invested in estimating the rate at which the LSST will discover interstellar objects, which has been most recently estimated at 1-3 objects per year \citep{Hoover2022}. \citet{Flekkoy2023} concluded that the LSST should detect a second `Oumuamua-like interstellar object in $<5$ yr after the survey starts with  90$\%$ confidence.  Previously, \citet{2019ApJ...884L..22R} concluded that LSST should be able to detect over 100 objects per year with radii larger than 1 meter. These estimates naturally contrast with those provided prior to the discovery of 1I/`Oumuamua and 2I/Borisov \citep{2009ApJ...704..733M}, which projected a low probability for LSST to detect an ISO during its operational lifetime.

\begin{figure*}
\begin{center}
\includegraphics[width=\linewidth]{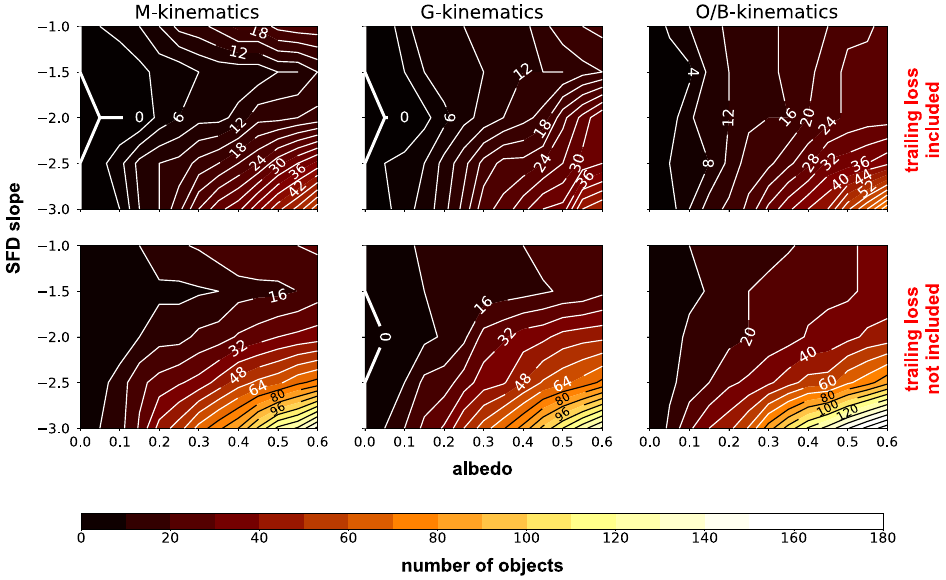}
\caption{The total number of detectable interstellar objects per year. The upper and lower panels show this number when the effects of trailing loss are included and excluded. The  detectability criterion requires at least 3 detections in the synthetic LSST frames.  The trailing loss depends on the exposure time and FWHM. Here we adopt a 0.7 arc seconds for FWHM and 30 s exposure time, as given in \cite{2018Icar..303..181J}.   We cut the albedo axis at 0.6  because the number of asteroids expected beyond this value is considered negligible \citep{Mainzer2012}.}\label{Fig:number_objects_trailingloss}
\end{center}
\end{figure*}

\begin{figure}
\begin{center}
\includegraphics[width=\linewidth]{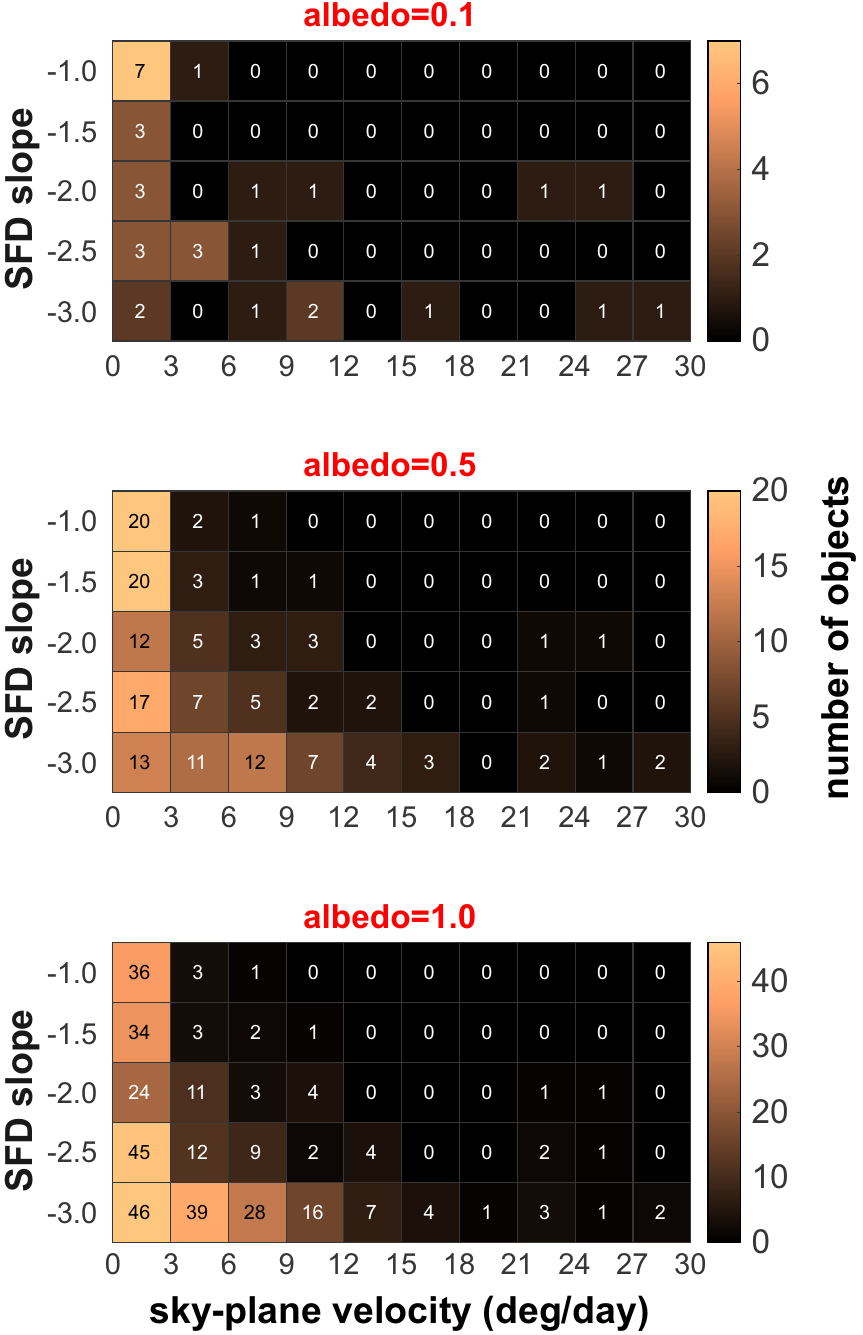}
\caption{The apparent rate of sky motion of interstellar objects as a function of the SFD-slope. The three panels correspond to three assumed albedos for each object in the synthetic populations.    The rates of motion are orders of magnitude higher than for typical solar system objects.}\label{Fig:apparentmotion}
\end{center}
\end{figure}

\begin{figure}
\begin{center}
\includegraphics[width=\linewidth]{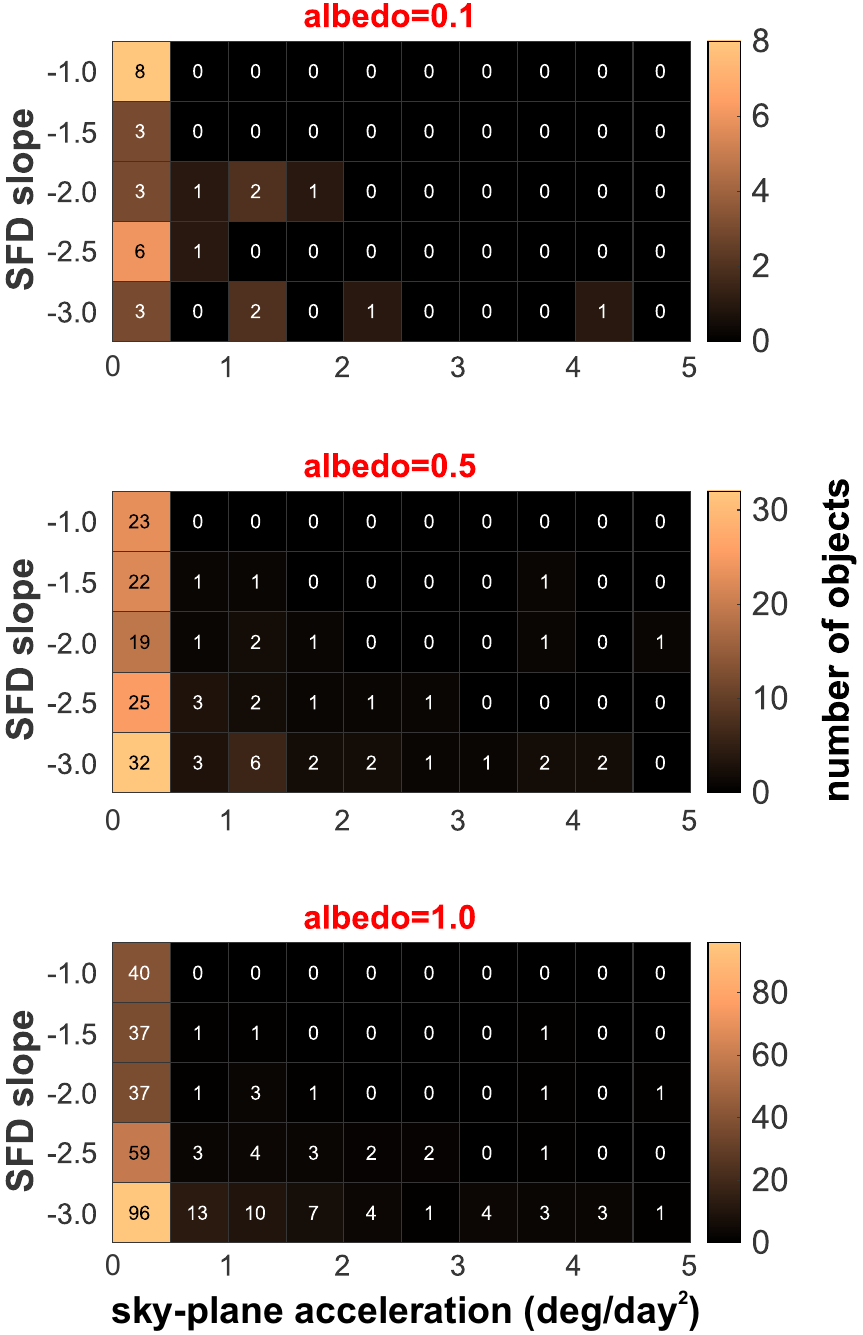}
\caption{Same as Figure \ref{Fig:apparentmotion} but for the acceleration of the sky-motion. We only show the absolute value of the apparent acceleration because the sign does not impact  the detectability. }\label{Fig:apparentacceleration}
\end{center}
\end{figure}

\begin{figure}
\begin{center}
\includegraphics[width=\linewidth]{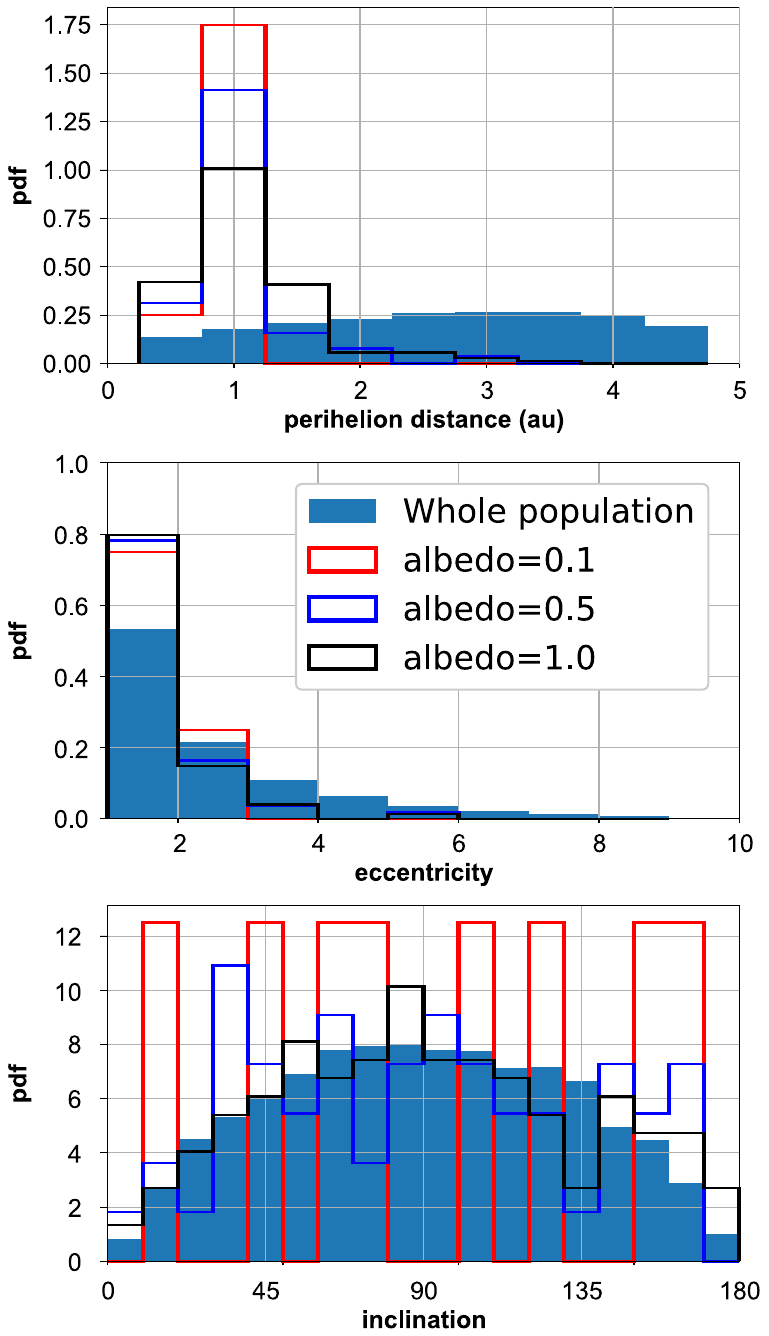}
\caption{The perihelion distance (top), eccentricity (middle) and inclination (bottom) of detectable interstellar objects. The underlying population has a SFD slope of -3 and  each object has an albedo of 0.1. We show the distributions for the entire population without the detectability constraints (blue solid), and for cases where the albedo is 0.1 (red unfilled), 0.5 (blue unfilled) and 1.0 (black unfilled).}\label{Fig:orbitalelements}
\end{center}
\end{figure}

The most straightforward method to calculate  the distribution of expected trajectories is via a  Monte Carlo integration, defined as the ``dynamical method"  \citep{Marceta2023} and implemented in \citet{Cook2016}, \cite{Engelhardt2014}, \cite{Seligman2018} and \cite{Hoover2022}. Alternatively, the analytic ``probabilistic method" is orders of magnitude more computationally efficient \citep{Marceta2023}. In this paper, we implement the probabilistic method in conjunction with the objectsInField (OIF) \citep{Cornwall2020}  software to estimate  the population of interstellar objects that will be discovered with the LSST. Given the computational efficiency of the method, we are able to produce synthetic survey results for   a range of assumed  SFDs, albedos, and kinematics.

This paper is organized as follows: In \S \ref{sec:methods}, we describe the methodology of our study, primarily focusing on an overview of the previously reported probabilistic method \citep{Marceta2023} and our implementation of the LSST observability criteria. In \S \ref{sec:results} we present the results of the distribution of interstellar objects that will be detectable by LSST as a function of the assumed SFD slope, albedo and background stellar kinematics of the galactic population. In \S  \ref{sec:conclusions} we conclude.
 
\section{ Methods} \label{sec:methods}

\subsection{Overview of Probablistic Method}

The method that we use to generate the population is briefly described in this section. The probabilistic method  calculates the distributions of orbital elements of interstellar objects as a function of the assumed kinematic distribution. It is orders of magnitude more computationally efficient than the dynamic method that has previously been implemented by \citet{Cook2016,Engelhardt2014,Seligman2018,Hoover2022}. Specifically,  starting from the assumption of conservation of the total number of interstellar objects in arbitrary sphere around the Sun, it derives the joint probability density functions of 6 parameters which define orbits of interstellar objects in the form:

\begin{equation}\label{eq:orbelems}
    \begin{aligned}
&p=\\
&\begin{cases}
\displaystyle
\frac{v_{\infty}p }{4\pi r}\frac{B}{\sqrt{v_{\infty}^2r^2+2\mu r-B^2v_{\infty}^2}}, & \text{$B \leq B_1$}\\\\\\
\displaystyle
\frac{v_{\infty} p}{2\pi r}\frac{B}{\sqrt{v_{\infty}^2r^2+2\mu r-B^2v_{\infty}^2}}, & \text{$B_1< B \leq B_2$}
\end{cases}
\end{aligned}
\end{equation}
where $\mu$ is gravitational parameter of the Sun, $v_{\infty}$ is excess velocity, $r$ is heliocentric distance, $B$ is impact parameter, $B_1$ and $B_2$ are critical impact parameters defining if an object hits the Sun ($B<B_1$) or misses the heliocentric sphere with radius $r$ ($B>B_2$), and   $p$ represents the kinematics of the population in interstellar space (i.e distribution of interstellar velocity components with respect to the Local Standard of Rest).

The probabilistic method incorporates   the gravitational focusing from the Sun by construction (see Section 2 in \citet{Marceta2023}). This method can incorporate any assumed kinematic distribution of interstellar objects. In this paper, we simulate interstellar objects assuming that they exhibit the same   kinematics as  M-, G- and O/B-star populations \citep[see the discussion in][]{Seligman2018,Hoover2022,Marceta2023}.

\subsection{LSST Observability Criteria: objectsInField Implementation}

\citet{Cornwall2020} presented an open source software objectsInField (OIF). This software simulates a realistic LSST campaign that incorporates observation scheduling.  Specifically, we utilized the cadence labeled  {\emph {kraken\_2026}} which is widely considered as a top candidate for the observational baseline cadence. This and other cadences are described in detail in the \href{https://docushare.lsst.org/docushare/dsweb/Get/Document-28716}{Alternate Observing Strategies}. Certain alternative strategies involve only minor modifications to {\emph {kraken\_2026}} (e.g. {\emph {colossus\_2665}}) and are not expected to significantly impact the performance for ISO detections. However, some other proposed cadences employ significantly different strategies such as the rolling cadence (e.g. {\emph {kraken\_2036}}). These alternative strategies focus on a single region of the sky at a time instead of spreading observations across the entire visible sky every few days. Performance estimation of these strategies would therefore  require a separate systematic analysis for ISOs and solar system objects \citep{2023ApJS..266...22S}. 

The OIF software uses a synthetic Solar System model developed by \citet{Grav2011}. This package generates a list of candidate detections for an input population of moving objects in a specified list of field pointings. This tool only provides objects that will appear in each specific field of view (FOV). There will be about 2.4 million FOVs over 10 years of the survey. However, the software does not assess the detectability of the object and identify which trajectories would realistically be detected. Therefore, our nominal detections are simply objects that will appear bright enough in a realistic LSST FOV.  We transformed diameters to absolute magnitudes according to the conversion formula \citep[see e.g.][]{bowell-etal_1989, 2007Icar..190..250P}

\begin{equation}
   H = 15.618 - 2.5 \log_{10}(albedo) - 5 \log_{10}(D),
   \label{eq:mag2dia}
\end{equation}
 where the diameter, $D$, has units of  kilometers. An apparent visual magnitude is than calculated by OIF as

\begin{equation}
   V = H + 5\log_{10}(\Delta) + 5\log_{10}(r_h) - \Phi.
   \label{eq:magvis}
\end{equation}
 In Equation \ref{eq:magvis}, the phase function, $\Phi$, is calculated according to

\begin{equation}\label{eq:phase_function}
    \begin{aligned}
&\Phi=2.5\log_{10}((1-G)\cdot \varphi_1 + G \cdot \varphi_2)\,,\\
&\varphi_1=\exp{\left(-A_1\tan^{B_1}\left(\phi \over 2\right)\right)}\,,\\
&\varphi_2=\exp{\left(-A_2\tan^{B_2}\left(\phi \over 2\right)\right)}\,,
\end{aligned}
\end{equation}
where $A_1=3.33$, $A_2=1.87$, $B_1=0.63$, $B_2=1.22$, $G=0.15$ and $\phi$ is a phase angle.

 We consider an object detected if it reaches a visual magnitude of 24.38, which corresponds to the  expected  average $5\sigma$ limiting magnitude in the \emph{g} filter. Furthermore, we require that   objects  have \textit{at least 3} detections to be included in the detected population. A linking algorithm could optimistically connect 3 trailed detections into a preliminary orbit. The rate of transforming these detections into discoveries relies on the linking algorithms' capacity to connect detections with a range of relevant features. For example, the algorithm would need to overcome   ISO sky-plane velocities and accelerations that are   much more rapid than those of any solar system populations.

\begin{figure}
\begin{center}
\includegraphics[width=\linewidth]{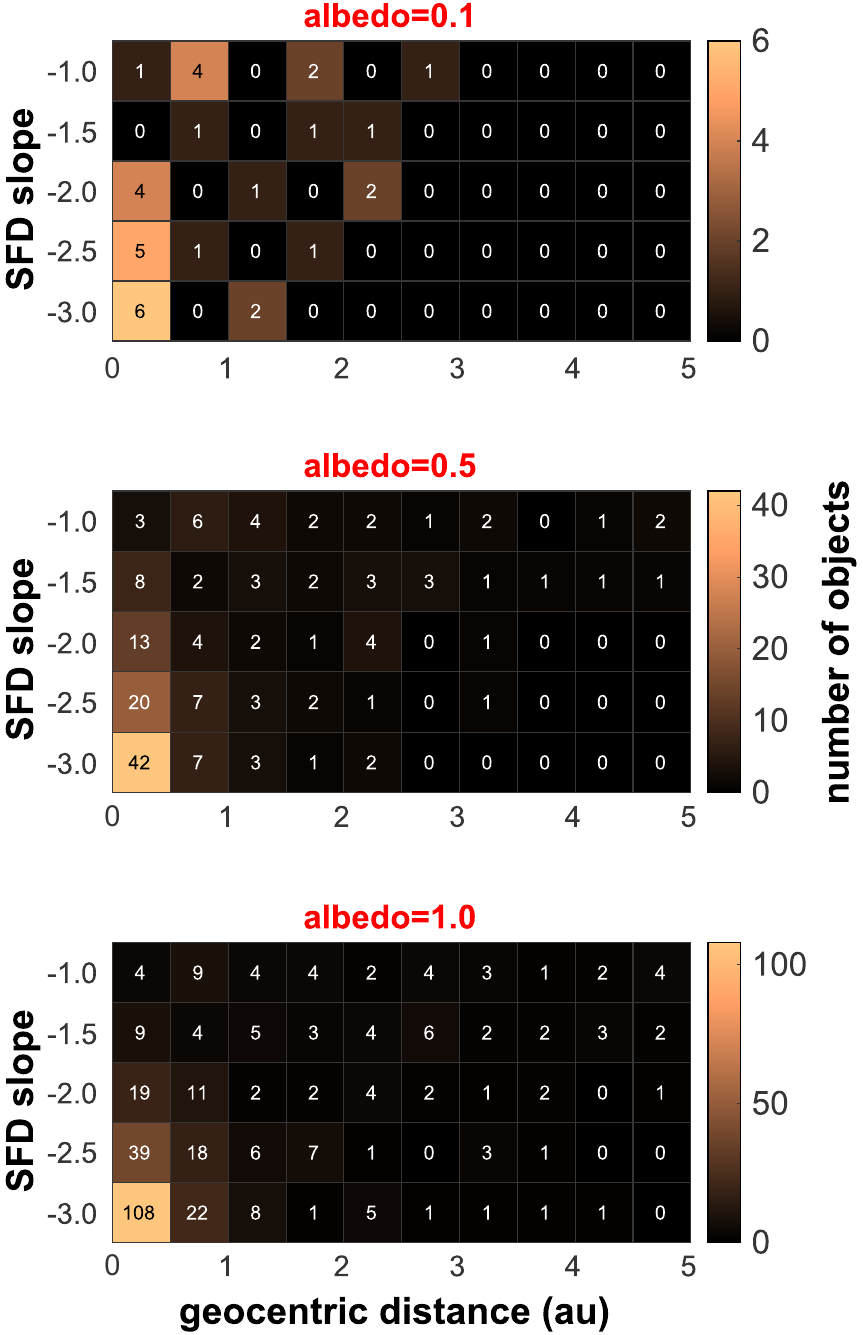}
\caption{The distribution of geocentric distance of detectable interstellar objects when they attain maximum brightness.}\label{Fig:mediangeocentricdistance}
\end{center}
\end{figure}

\begin{figure}
\begin{center}
\includegraphics[width=\linewidth]{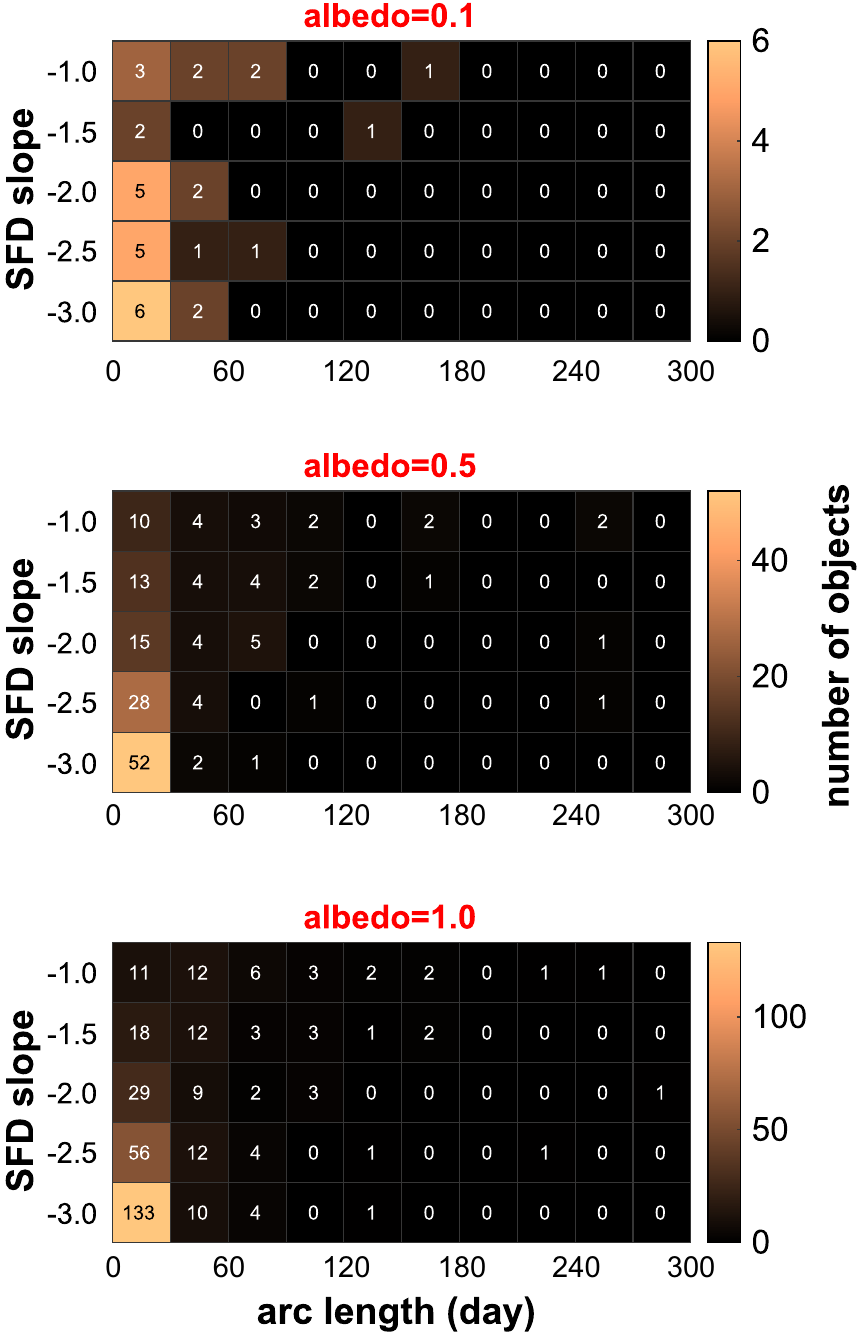}
\caption{The distribution of arc length during which a synthetic interstellar object is detectable. }\label{Fig:arclengthdistribution}
\end{center}
\end{figure}

\begin{figure}
\begin{center}
\includegraphics[width=\linewidth]{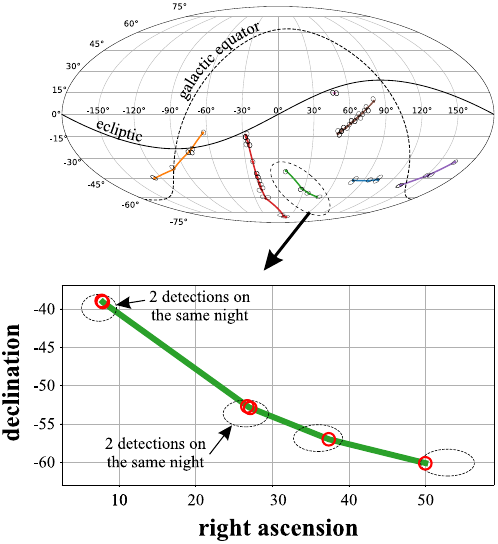}
\caption{Examples of sky paths of synthetic detectable objects (top panel). The individual points show individual detections with the OIF simulation.  We show one zoomed in region (indicated in the top panel in the dashed circular region) including a synthetic interstellar object path with 6 detections (lower panel).  These plots are obtained assuming a SFD of -2.5 (an intermediate value),  albedo of 0.1 (conservative) and O/B kinematics.  }\label{Fig:skypath}
\end{center}
\end{figure}

\begin{figure}
\begin{center}
\includegraphics[width=\linewidth]{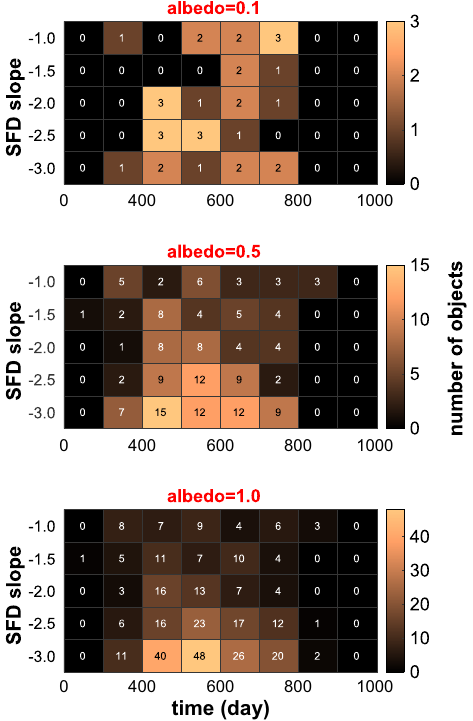}
\caption{The amount of time a detectable interstellar object spends inside of the 5 au sphere for a range of albedo and SFD slope. }\label{Fig:timespent5aub}
\end{center}
\end{figure}

\begin{figure}
\begin{center}
\includegraphics[width=\linewidth]{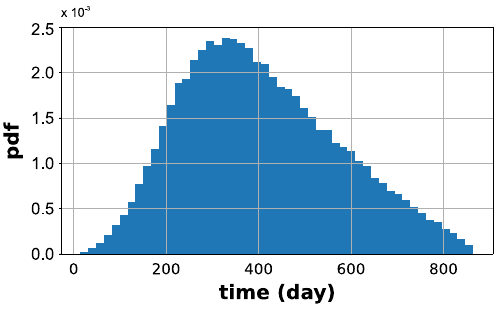}
\caption{The amount of time an interstellar object spends inside of the 5 au sphere for O/B kinematics. This distribution is for the entire population, and is not filtered by detectable objects. }\label{Fig:timespent5au}
\end{center}
\end{figure}

\begin{figure}
\begin{center}
\includegraphics[width=\linewidth]{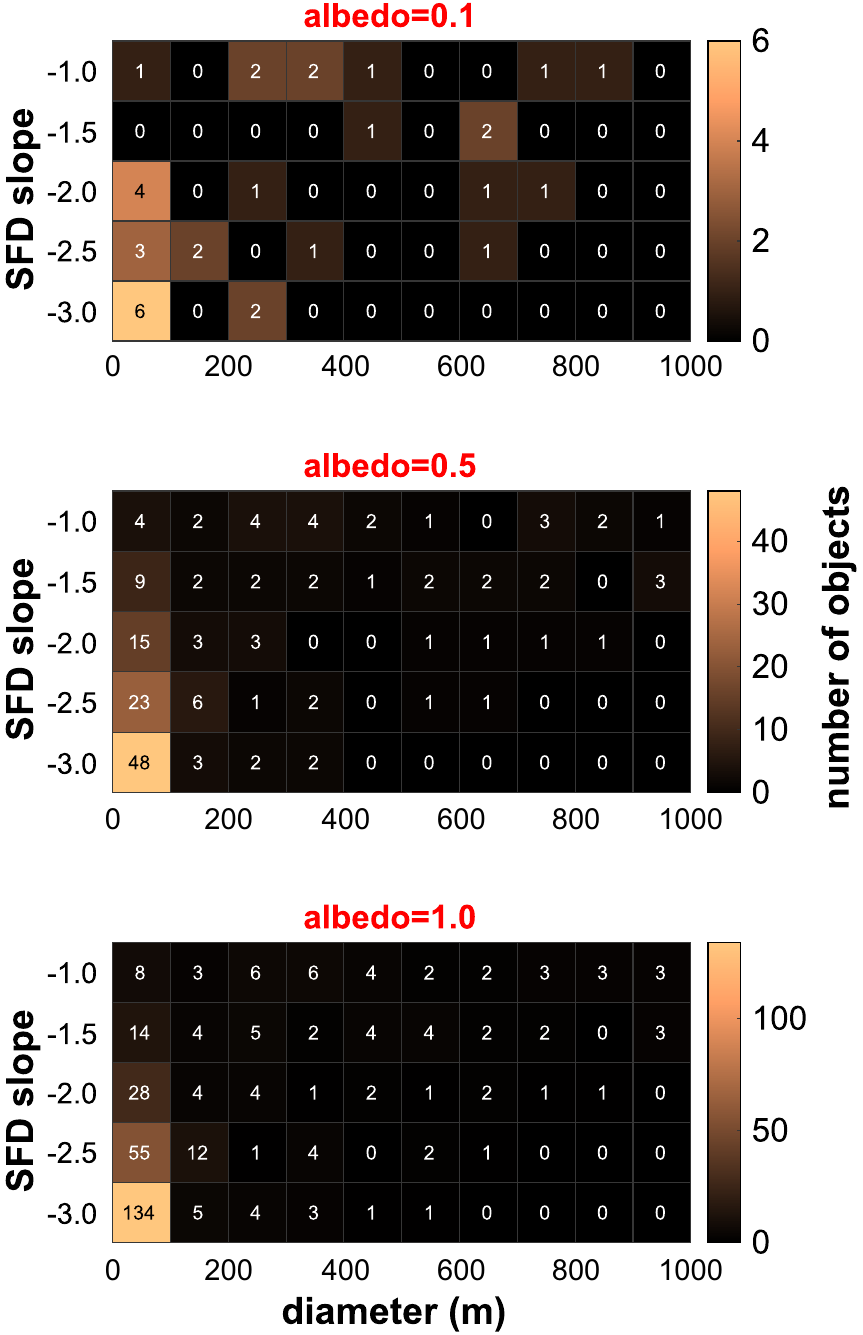}
\caption{The distribution of the diameters of synthetic interstellar objects detectable with the LSST. }\label{Fig:diameters}
\end{center}
\end{figure}

\section{Results}\label{sec:results}

In this section, we describe the results of our synthetic population synthesis. We initiate the population at a heliocentric distance of 30 au. This is the maximum \textit{initial} heliocentric distance from which the brightest object (D=1 km, albedo=1) can reach a  distance where it can be observed ($\sim$8 au) in a 1 year simulation time.
We assume that the population begins with every object having an albedo of 1.0, and we then scale the analysis for a variety of albedos. 

\subsection{Number of Interstellar Objects Detectable with LSST}

We generate synthetic populations for a range of assumed SFDs. We simulated objects assuming that the sizes range from 10 m to 1000 m, although the results are not sensitive to the exact cutoff size. Objects larger than 1 km are very rare in general. On the other hand, objects smaller than 10 meters are very faint and rapidly moving. All of these  objects are normalized to the spatial number density  of 0.1 au$^{-3}$ for objects larger than 100 meters in diameter, implied by the detection of 1I/`Oumuamua for similar sized bodies. 

 We restrict our analysis to asteroidal ISOs. A similar analysis will be conducted incorporating  a variety of cometary brightening models to estimate the total number of detectable active interstellar comets. This analysis will also assess the extent to which selection effects impact the ratio of cometary to asteroidal detections and how this reflects  the properties of the underlying true populations.

In Figure \ref{Fig:number_objects_trailingloss}, we show the number of interstellar object detections per year with the LSST as a function of assumed albedo and size-frequency distribution. The criterion for detection is that the object is detected at least 3 times. These detections  may occur on the same or on different nights. This could prove to be a crucial issue based on the linking algorithms used on the LSST data \citep{2018Icar..303..181J, 2018AJ....156..135H}. This analysis can be re-performed using different detection criteria to modify the synthetic survey results.  

We use the objectsInField (OIF) algorithm to simulate the synthetic detections. We perform the simulations for a range of assumed albedo and SFDs (described by a power law $N\left(D\right) \propto D^q $), where the albedo ranges from 0 -- 1 and the SFD slope $q$ ranges from -3 to -1. This is intended to cover a substantial portion of possible realistic populations. We also show the same plot assuming (i) M- (ii) G- and (iii) O/B- kinematics for the kinematic distributions of interstellar objects. It is apparent that the number of detectable objects with LSST varies between $\sim 0-70$ per year based on the underlying kinematics, SFD slope and albedo of the population,  as shown in the upper panels of Figure 1 which include the effect of trailing loss.

 We account for trailing loss in our synthetic detections which is especially important for rapidly moving, inactive interstellar objects. Trailing loss occurs when an object's motion causes its photons to spread across a broader area than a typical stellar point spread function. This effect is quantified by employing the function \citep{2018Icar..303..181J}

\begin{equation}\label{eq:trailling_loss}
\begin{aligned}
& \Delta m = -1.25 \log_{10} \left(1+\frac{ax^2}{1+bx}\right),\\\\
& x=\frac{v T_{exp}}{24 \theta},
\end{aligned}
\end{equation}
where $v$ is the sky-plane velocity (in deg/day), $T_{exp}$ is the exposure time (in seconds), and $\theta$ is the full-width half-maximum (FWHM) (in arcseconds). \citet{2018Icar..303..181J} found that the parameter values $a=0.67$ and $b=1.16$ best describe trailing SNR losses. We employ $\theta=0.7$, which is expected to represent the median seeing (\href{https://rtn-022.lsst.io/}{Seeing values for LSST strategy simulations}), and $T_{exp}=30 , s$, corresponding to the {\emph{kraken\_2026}} cadence.

In the lower panel of Figure  \ref{Fig:number_objects_trailingloss} we  show the  number of detected objects  \textit{without} including the trailing loss in the calculation. It is evident that the number of detectable objects increases by a factor of $\sim3-4$ when the trailing loss is not incorporated. The reason for this is that the interstellar objects tend to travel extremely fast and have high apparent rates of sky motion.  This proportion grows as the SFD slope increases and the characteristic albedo decreases. However, the only scenarios resulting in 0 detections are those with (i) an albedo below 0.1, (ii) an SFD slope of -2, and (iii) M- and G-kinematics when the trailing loss is included. This analysis strongly supports the conclusion that, based on the assumed number density, LSST will consistently detect ISOs. In addition, it is possible that the number densities inferred for this population from the detections of 1I/`Oumuamua and 2I/Borisov are under representing the true number density, simply because the population of objects move too rapidly and are undetectable due to trailing loss. Moreover, if techniques are developed to detect objects that are very rapidly moving then our estimates for the detection would increase. 

For the remainder of this work, all other results are shown using only O/B kinematics. This choice is motivated because , for a given number density, these younger stellar populations yield the largest number of objects, yielding the most computationally efficient statistics. This is due to the fact that these young stars have smaller velocity dispersions, and therefore the solar gravitational focusing is more efficient and produces a larger number of objects close to the Sun. Given that there are already large number of parameters,  we elected to not show the remaining results for all three kinematics.  However, the kinematic distribution of ISOs is entirely unconstrained, as is the relative contribution of different star populations. Furthermore, \citet{2023arXiv230805801H} suggests that the ISO population should be drawn not only from the current stellar populations but from a so-called \emph{sin morte} population of stars, encompassing all stars since the birth of the Galaxy. Once the kinematic distribution of interstellar objects is better constrained via future detections, these results can be updated.

\subsection{Sky Motion Statistics}
In Figure \ref{Fig:apparentmotion} and Figure \ref{Fig:apparentacceleration} we show the apparent rate of motion and acceleration of detectable objects as a function of the SFD slope and albedo. It is evident that there are many objects that cannot be tracked by traditional algorithms because they are moving much faster than the proposed limits for apparent motion of $0.5^\circ d^{-1}$ \citep{2018Icar..303..181J}. This  also applies to the apparent acceleration.  

 An increase in the SFD slope  generally leads to an increase in the number of  rapidly moving objects. This is because the smaller objects must exhibit closer approaches to the Earth in order to be detectable, and they are therefore moving much more rapidly.  A non-trivial  corollary to this is that the distribution of sky motions of interstellar objects provides information regarding the SFD.
 
\subsection{Trajectories of Interstellar Objects}
In Figure \ref{Fig:orbitalelements} we show the  perihelion distance, eccentricity and inclination  for the detected populations. By renormalizing we also show the corresponding distributions for the entire  population. The different color histograms correspond to  detectable objects with albedos of 0.1, 0.5 and 1. The eccentricity distribution is independent of albedo, but the peak for perihelion distance at 1 au increases when the albedo decreases.  The inclination distribution converges to the underlying distribution of inclination for the higher albedo cases. However there are only 8 objects  in the low albedo case  which is an insufficient sample size for statistical measurement.

In Figure \ref{Fig:mediangeocentricdistance} we show the distribution of the geocentric distance of objects when they are at maximal brightness. These distributions are straightforward to interpret. In the case of a low assumed albedo for the population, the detections are mostly at small geocentric distances. When a larger albedo is assumed, there are some detections at larger distances. These more distant objects most likely have smaller rates of motion as well.

In Figure \ref{Fig:arclengthdistribution} we show the arc length for which an object is detectable for  each detectable object in the  populations. A smaller albedo assumed for the population produces  shorter arcs in general. The short arcs also dominate the high-albedo cases due to the large number of small objects in each population.  In general, objects are detectable for $<1-2$ months.

In Figure \ref{Fig:skypath} we show examples of sky paths of detected synthetic objects on the sky plane. The individual points show individual detections with the OIF simulation. These plots are obtained for SFD -2.5,  albedo = 0.1 and O/B kinematics. We also required that the object be within the LSST FOV for every synthetic detection. Synthetic interstellar objects are typically moving rapidly and often do not appear in the same FOV twice during one night. As evident from the figure, the typical paths and detection patterns are  diverse.

In Figure \ref{Fig:timespent5aub} and \ref{Fig:timespent5au} we show the duration of time that interstellar objects spend within  5 au. We show this for the detectable population (Figure \ref{Fig:timespent5aub}) and the entire population (Figure \ref{Fig:timespent5au}). This parameter is a critical component in calculating detection rates, because it is used in conjunction with the number density to estimate rates. The time spent within the 5 au sphere is relatively insensitive to the albedo and SFD. Moreover, the median is $\sim1-2$ yr for all populations.  The distribution in Figure \ref{Fig:timespent5au} is skewed toward shorter times. This is because there are relatively more objects with large perihelion distances. These objects  only barely pass within the 5 au sphere. However, these distant objects are never detectable. Detectable objects, on the other hand, have smaller perihelion distances with more curved orbits, and spend significantly more time inside the sphere.

\subsection{Sizes}

In Figure \ref{Fig:diameters} we show the distribution of sizes of detectable objects for a range of SFD and albedo.  Populations with low values of the SFD  slopes produce a relatively uniform distribution of sizes of detectable objects. On the other hand, populations with larger SFD slopes produce many more \textit{small} detectable objects. This is also the case for the entire population. In general,  interstellar objects in the 100 meter  size range are more likely to be detected  for a steeper SFD. We conclude that given that 1I/`Oumuamua was the first object discovered, this is tentative evidence for a steeper SFD of interstellar objects. We note  that this conjecture is highly speculative, and the inherent SFD will be constrained with future detections.

\begin{figure}
\begin{center}
\includegraphics[width=\linewidth]{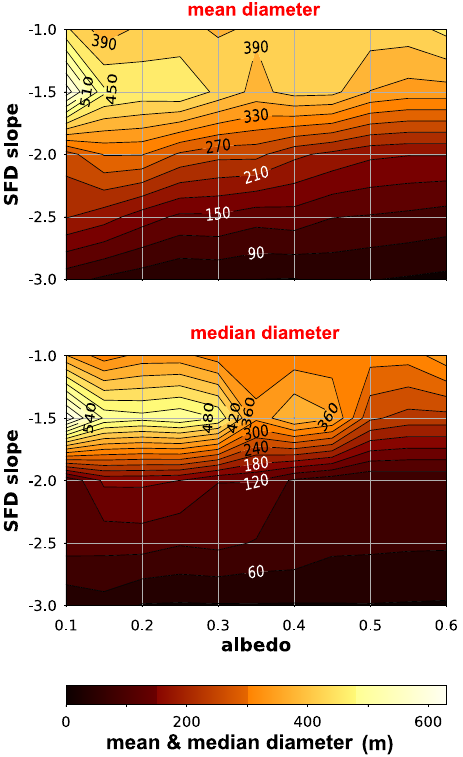}
\caption{Mean and median diameter of detectable objects. }\label{Fig:mediandimater}
\end{center}
\end{figure}

In Figures \ref{Fig:mediandimater} we show the median and mean diameters of detectable objects as a function of albedo and SFD slope. The typical sizes of interstellar objects detected in the future with the LSST will be between 50-600 meters based on the SFD and albedo. It appears that the diameter of detectable interstellar objects is relatively sensitive to the SFD and insensitive to the albedo. Moreover, it is possible that the detectability of smaller objects is more heavily effected by the trailing loss (Figure \ref{Fig:number_objects_trailingloss}) than that of larger objects. This may contribute to the lack of dections of smaller interstellar objects, although future work is required to quantify this effect.

\section{Discussion}\label{sec:conclusions}

In this paper, we generated a synthetic population of interstellar objects following the methodology developed in \citet{Marceta2023}, assuming different kinematics and a range of size-frequency distributions. We then implemented the Objects in Field algorithm to generate realistic LSST survey conditions and evaluated the interstellar objects that would be detectable with the LSST. We required for detectability that the objects appeared bright enough in the LSST frames and produced at least three detections, regardless of the nights that they were detected in. We also accounted for the trialling loss due to  large sky-plane velocities of interstellar objects.

We found that the overall rate at which the LSST discovers interstellar objects is  sensitive to the size-frequency distribution slope and albedo of interstellar objects. The annual rate at which LSST should discover 'Oumuamua-like interstellar objects ranges from $\sim 0-70$ detected objects per year based on the assumptions of the SFD and albedo.  The span of this range is considerably broader than some of the previous estimates. For example, \citet{Hoover2022} predicted that the LSST would make 1-3 ISO detections per year. However, those simulations did not incorporate the size–frequency distribution of interstellar objects. Here   we have demonstrated that changes in the SFD can  substantial increase the number of detectable interstellar objects and potential targets for rendezvous missions. Additionally, \citet{Hoover2022} assumed that all objects exhibit the same absolute magnitude as 1I/`Oumuamua (22.4). Given that our analysis spans a broad range of SFDs and characteristic albedo values, it has yielded a wider spectrum of potential detection numbers. This bracketed range covers a significant range of solar system populations at small sizes \citep{Grav2011, 2020AJ....159..148P}, so we can safely assume that the number per year will be within this range. 

In this work, we presented results that were intentionally agnostic to the interstellar object SFD and albedo distribution. This is due to the fact that these parameters are almost entirely unconstrained empirically. Moreover, measurements of the SFDs within and across various small body populations within the solar system vary and are size limited. If the albedo distribution of interstellar objects is similar to those measured in asteroids from the NASA’s Wide-field Infrared Explorer (WISE) mission  \citep[in the range of 0.01-0.4][]{Mainzer2012}, $0-15$ interstellar objects should be detectable per year with LSST.

The apparent rate of motion appears to be the largest issue to overcome for detecting interstellar objects with the LSST. In Figure \ref{Fig:apparentmotion} it is evident that these objects move very rapidly on the sky. Depending on SFD slope and albedo, the fraction of objects faster than 1 $^{\circ}d^{-1}$ ranges from 16\% to 85\%. Increasing the SFD and decreasing the albedo increases this fraction. In order to link these detections into orbits, a suitable detection algorithms relying on the length and orientation of the trails will be necessary.  After  accounting for exposure time and expected seeing, it is evident that  objects moving faster than $\sim 1$ $^{\circ}d^{-1}$ will exhibit trailing  in LSST images \citep{2020Icar..33813517F}. Therefore, suitable algorithms for trailed detections should be employed to link these detections \citep{2017AJ....154...13V}. A potential challenge lies in objects with velocities ranging between 0.5 and 1 $^{\circ}d^{-1}$. While these objects will not create trails, they exceed the typical suggested velocity threshold of 0.5 $^{\circ}d^{-1}$ (\href{https://www.google.com/url?sa=t&rct=j&q=&esrc=s&source=web&cd=&ved=2ahUKEwiGqM-47YSBAxXDi_0HHX0pDPMQFnoECBYQAQ&url=https%3A%2F%2Fdocushare.lsstcorp.org%2Fdocushare%2Fdsweb%2FGet%2FRendition-27268%2Funknown&usg=AOvVaw1Q2Gz9gO12CS2xiyEYV49m&opi=89978449}{LSST Moving Object Pipeline System Design}). The fraction of our simulated detections falling within this range varies from 7\% to 55\%, depending on the kinematics, SFD, and albedos of the population. This strongly suggests the need to increase the velocity threshold in linking algorithms cover these detections. The use of more efficient algorithms, such as HelioLinC \citep{2018AJ....156..135H}, has the potential to facilitate  this threshold increase  and to handle a larger number of tracklets.
 
An intriguing suggestion from these results is that it may be easier to identify rapidly moving objects outside of the densely populated ecliptic plane. For example, the Antarctic Search for Transiting Exoplanets (ASTEP) project field of view from Antarctica could be ideal for finding interstellar objects \citep{hasler2023small}.  While interstellar objects will also populate the ecliptic plane, they will be more or less isotropically distributed across the sky for surveys with limiting magnitudes comparable to the LSST. However, detections of interstellar objects like 1I/`Oumuamua cluster around the ecliptic for less sensitive surveys  \citep{Hoover2022}.

It appears that the trailing loss is another major  limiting factor on the number of detectable interstellar objects with the LSST (Figure \ref{Fig:number_objects_trailingloss}). It is possible that the number density of `Oumuamua-like objects is higher than  currently  estimated due to a large fraction of interstellar objects currently undetectable due to trailing loss and rapid sky motions.  To some extent, ISO discovery is similar to NEO discovery. The primary differences are  that the spatial distribution of ISOs is isotropic and they exhibit a broader range of sky-plane velocity and acceleration. The development of  linking algorithms optimized for rapidly moving objects with trailing loss would be ideal for detecting interstellar objects.


 \section{Acknowledgments}
 We thank Ari Heinze, Dave Jewitt, Dong Lai, Matt Payne and Matt Holman for useful conversations and suggestions.  D.Z.S. acknowledges financial support from the National Science Foundation  Grant No. AST-2107796, NASA Grant No. 80NSSC19K0444 and NASA Contract  NNX17AL71A. D.Z.S. is supported by an NSF Astronomy and Astrophysics Postdoctoral Fellowship under award AST-2202135. This research award is partially funded by a generous gift of Charles Simonyi to the NSF Division of Astronomical Sciences.  The award is made in recognition of significant contributions to Rubin Observatory’s Legacy Survey of Space and Time.  DM acknowledges financial support from Ministry of Science, Technological Development and Innovation of the Republic of Serbia, contract No.  451-03-47/2023-01/200104.

\bibliography{bibliography}{}
\bibliographystyle{aasjournal}
\end{document}